# Impact Angle Influence in High Velocity Dust Collisions during Planetesimal Formation

Jens Teiser*, Markus Küpper, Gerhard Wurm

Faculty of Physics, University Duisburg-Essen, Lotharstr. 1, D-47058, Germany

Corresponding author: jens.teiser@uni-due.de, phone: 0049 203 379 2941

markus.kuepper@stud.uni-due.de, gerhard.wurm@uni-due.de

**Abstract:**

We have examined the influence of impact angle in collisions between small dust aggregates and larger dust targets through laboratory experiments. Targets consisted of µm-sized quartz dust and had a porosity of about 67%; the projectiles, between 1 and 5 mm in diameter, were slightly more compact (64% porosity). The collision velocity was centered at 20 m/s and impact angles range from 0° to 45°. At a given impact angle, the target gained mass for projectiles smaller than a threshold size, which decreases with increasing angle from about 3 mm to 1 mm. The fact that growth is possible up to the largest angles studied supports the idea of planetesimal formation by sweep-up of small dust aggregates.

**Introduction**

Coagulation processes are an important part of planet formation. The first growth steps are the coagulation of micrometer sized dust grains into larger dust agglomerates. During this phase the interactions between solid particles are determined by cohesion forces between single grains. Relative velocities between dust grains initially are very moderate, on the order of mm/s to cm/s, depending on the particle size (Weidenschilling & Cuzzi 1993). Several experimental and theoretical studies investigated this parameter range and showed that particles colliding at these velocities stick to each other once they touch (Blum & Wurm 2008 for review; Dominik & Tielens 1997; Wada et al. 2009; Langkowski et al. 2008; Wurm and Blum 1998). Due to the efficient sticking of small grains, growth in this very first phase is inevitable and mm-sized particles form rapidly. Larger dust aggregates will also form by mutual collisions but they will become more compact (Teiser & Wurm 2009b; Weidling et al. 2009; Teiser et al. 2011; Blum and Wurm 2000). In general, relative velocities between colliding bodies increase with increasing aggregate size. Collisions with particles of different size eventually can reach values up to several tens of m/s (Weidenschilling & Cuzzi 1993; Brauer et al. 2008).

In experimental studies Wurm et al. (2005) and Teiser and Wurm (2009a) showed that fragmentation of the projectile occurs at the highest expected impact speeds. However, this dissipates a large amount of the kinetic energy and a fraction (tens of %) of the projectile mass remains, stuck to the target. This way, mass gain of a larger body by collisions with smaller ones is possible at least up 56 m/s (Teiser & Wurm 2009a). These experiments were limited to central collisions. The relation between impact angle and accretion has not yet been studied systematically, where we refer to the impact angle in this paper as the angle between the velocity vector of the incoming projectile (small aggregate) with respect to the surface normal of the target (large aggregate).

Evidence that mass gain is favored for small impact angles (central collisions) comes from experiments by Teiser and Wurm (2009b). They study how decimeter size bodies grow by accretion of small (100 µm) particles at 7.7 m/s. In these experiments Teiser and Wurm 2009b observed, that targets grow up to a maximum impact angle of 70°.

Assuming a spherical body moving through a stream of particles, the probability of impacts at a certain angle depends on the corresponding surface element exposed to the particle stream. For geometry reasons a large (e.g. decimeter) body in a protoplanetary disk accretes only a small fraction of its mass in central collisions. The most probable impact angle is then 45°. Therefore, net growth of a target through many collisions can only occur if mass gain is also possible in collisions at larger impact angles. We therefore present an experimental study on the influence of the impact angle on coagulation processes at large collision velocities.

**Experiments**

The experimental setup is described in Teiser and Wurm (2009a) and shown in Fig. 1 (a). The projectiles were accelerated by a crossbow launcher, giving a mean impact velocity of 19.7 m/s with a standard deviation of ± 1.0 m/s. Four projectiles were placed simultaneously in one projectile mount at the tip of the arrow. The arrow was stopped by a metal block with a central opening allowing the dust projectiles to pass, so only dust hit the target. The target was adjusted to impact angles between 0° (central impact) and 45° in steps of 5° each.

Projectiles and targets were prepared from the same broad size distribution of irregularly shaped grains of quartz . Grain sizes range from 0.1 µm to 10 µm, with 80 % of the mass between 1 µm and 5 µm. Blum et al. (2006) showed that the mechanical properties of dust agglomerates are mainly determined by the grain size distribution of the monomers. The detailed mineralogical and the chemical composition plays a minor role. The materials used within this study therefore can be treated as good analogues for silicate dust agglomerates in protoplanetary disks. The targets were produced by compressing dust in metal cylinders 10 cm in diameter by manually applying local pressure. Here, the surface element on which the pressure is applied is significantly smaller than the complete target. With this technique the degree of compression is smaller than for omni-directional compression, leading to a mean porosity of 0.67. The analysis of self-consistent growth of decimeter-size dust agglomerates by multiple impacts of small (100 µm) agglomerates showed that 68% porosity is a natural limit (Teiser & Wurm 2009a; Teiser et al. 2011), which is in good agreement with the samples prepared here. The projectiles were produced using a sieve with 0.5 mm or 1.0 mm mesh, depending on the projectile size to be used. This sieve was vibrated and round projectiles formed atop the mesh (not below). We measured the resulting projectiles to have a mean porosity of 0.64 with a standard deviation of ± 0.05. Weidling et al. (2009) studied the compaction of dust on a vibrated plate and found similar porosities of 0.64 ± 0.05.

The collisions took place in a vacuum chamber at ambient pressure below $10^{-1}$ mbar. They were observed with a camera at 500 frames $s^{-1}$ using stroboscopic illumination with a frequency of 1000 Hz. Impact velocities and projectile sizes were determined via image analysis by measuring the maximum diameters of the projectiles and their positions. The projectiles fragmented during the launching process in some experiments and, in such cases, the size of the largest fragment was used for the analysis. The projectiles used in this study were of the order of a few millimeters or smaller, as larger particles led to erosion at 20 m/s (Teiser & Wurm 2009a). The projectiles were small in comparison to the decimeter-size targets. Due to the large total mass of the target, the exact mass balance of the collisions could not be determined and we could not quantify an accretion or erosion efficiency. However, the outcome could well be classified qualitatively from target images after a collision in terms of mass loss or gain. Ejecta were collected to determine the mass distribution of the fragments, which was done by image analysis. Here, the given ejecta sizes were defined as the diameters of spheres with a corresponding projected area.

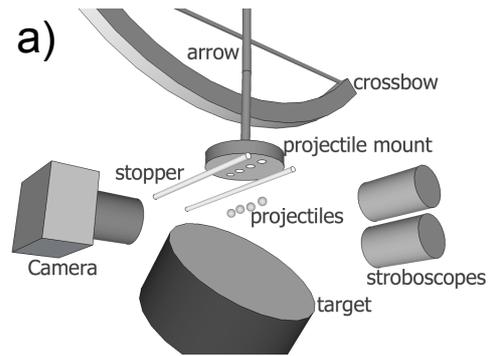
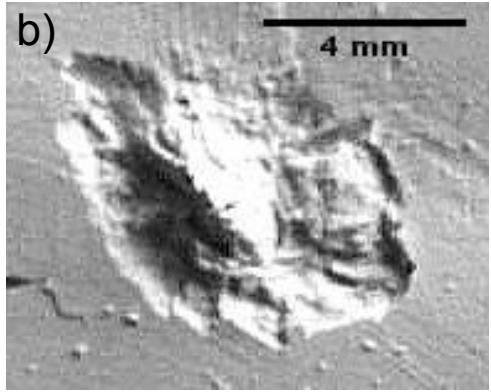
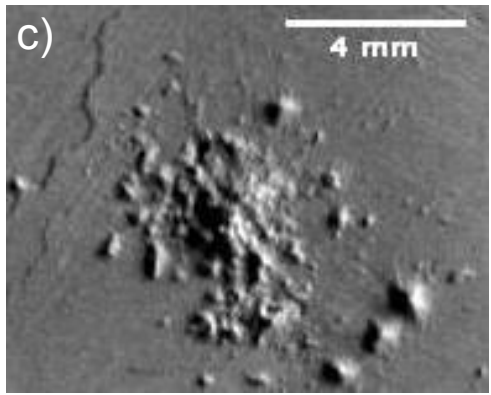
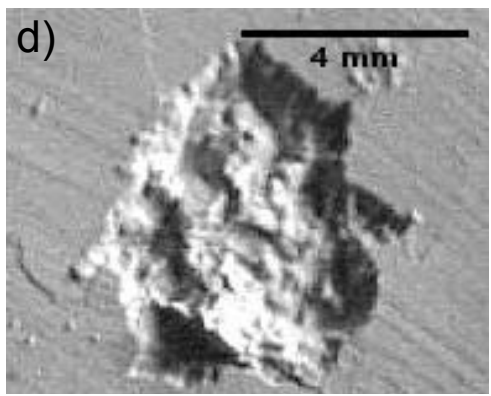

**Fig. 1:** Schematic sketch of the experimental setup (a) and examples for the three observed types of collision results. In panel b) target erosion is shown; a crater is the dominating feature. In panel c) target mass gain is shown; projectile material sticks; no crater is visible. Panel d) shows a combination of sticking projectile material and crater.

**Results**

Similar to the experiments on central collisions by Teiser and Wurm (2009a), three different collision outcomes have been observed, depending on the projectile size. Fig. 1 gives an example for the three different classes of collisions. The impacts can lead to erosion if the projectile craters the target and no significant amount of projectile material is left sticking to the target (Fig 1, b). Other impacts clearly lead to growth (mass gain) of the target, as projectile material sticks to the target surface and no crater occurs (Fig. 1, c). In between, as a third class, some impact results cannot unambiguously be claimed to be mass loss or gain. Here, projectile material sticks at the center of the impact site and is surrounded by a circular trench of ejected target material (Fig. 1, d). In case the balance between the eroded volume and the sticking material cannot be determined, these events are treated as neutral and are not considered further.

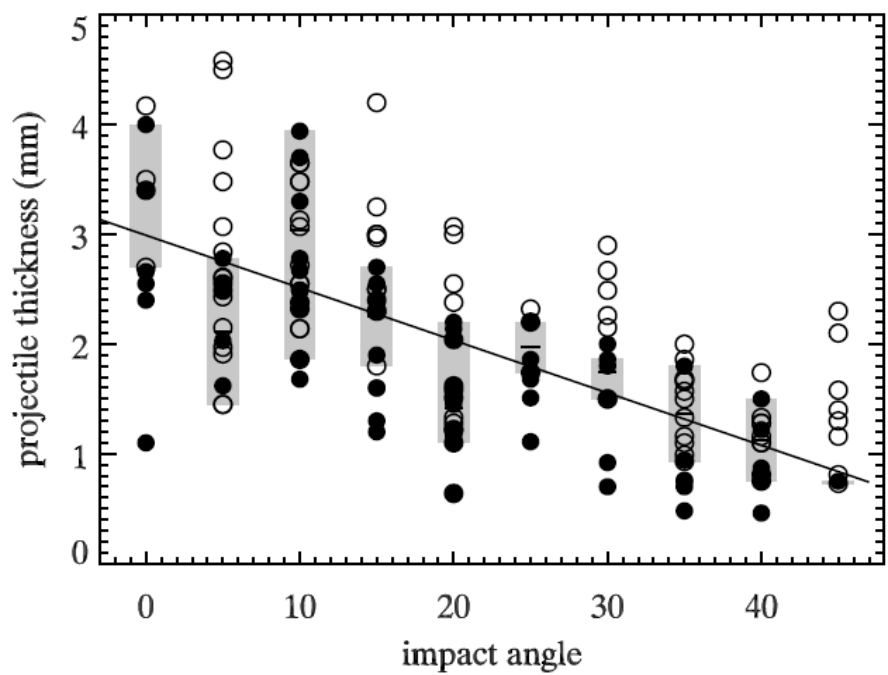

**Fig. 2:** Erosion and accumulation at different impact angles and projectile sizes (largest diameters in the camera images). Open circles mark erosion, filled symbols growth. The grey bars mark the transition regime between erosion and growth. Experiments without a clear mass balance are marked as horizontal black lines. The transition zones are defined as the size range between the largest projectile leading to growth (largest size, filled symbol) and the smallest projectile leading to erosion (smallest size, open circle). The solid line is a linear fit to the mean values of the transition zones (grey bars).

There is a clear relationship between the different collision outcomes and the projectile size. Fig. 2 shows the results of all collisions. The experiments show that there is a transition regime, in which collisions can lead to erosion or to growth. We define this transition regime as the size range between the smallest projectile size leading to erosion (open circle) and the largest projectile size leading to growth (filled symbol) for each impact angle. The transition regimes are marked by the grey bars in Fig. 2. We define the threshold projectile sizes, d, for erosion as the mean values of these transition regimes. In Fig. 2 a linear fit has been applied to the dependence of the threshold projectile size, d, on the impact angle, φ, giving

d = 2.962mm – 0.0476 (mm/°) · φ.                                                                    Eq. 1

At 0° the experiments reproduce the value for central collisions obtained by Teiser and Wurm (2009a) where the threshold between erosion and growth at impacts of 20 m/s was determined to be about 4 mm. The threshold size decreases with increasing impact angle to values smaller than 1 mm for impact angles of 45°, but growth is still possible even for impact angles as large as 45°.

During each impact a large number of fragments formed and left the impact site. Most of them were captured by a screen surrounding the target. Ejecta velocities are typically a few m/s or even lower. Experiments by Teiser and Wurm (2009b) showed that the size distribution of compact dust agglomerates is not changed significantly by impacts in this velocity range.

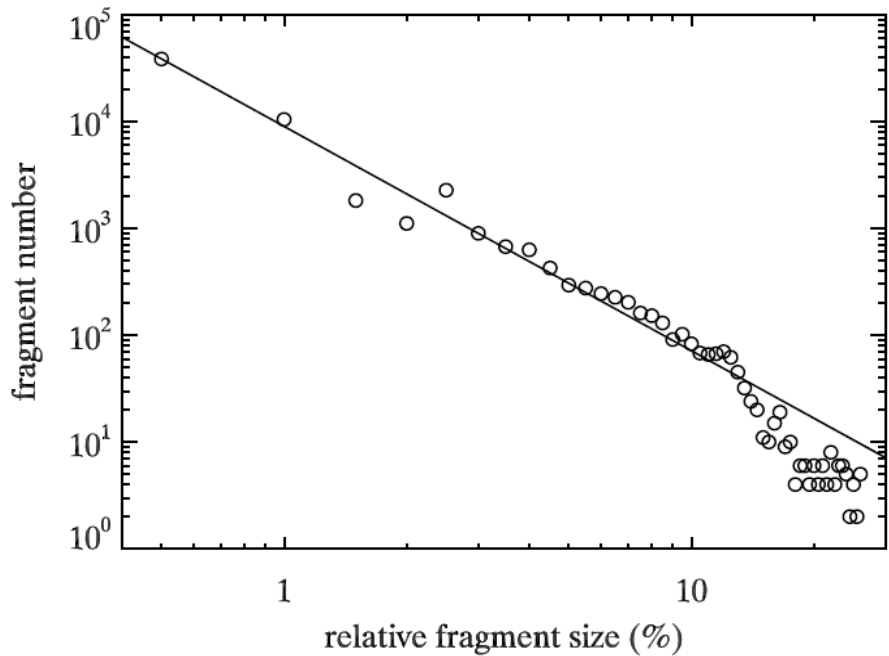

**Fig. 3:** Size distribution of the captured fragments as determined by image analysis. The sizes are diameters of spheres with equal projected surface and are normalized to the original projectile size.

The resulting fragment-size distribution does not depend on the impact angle but only on the size of the original projectile. To compare the results of different experiments and to improve the statistical significance, the fragment sizes are given relative to the size of the original projectile. Fig. 3 is therefore the sum of a number of size distributions, which were normalized by the projectile size before and are given as relative sizes to the original projectile size (in %).

The solid line is a power law with

$$n = 9000 \times s^{-2.1} \qquad \text{Eq. 2}$$

with the normalized projectile size s. The largest ejecta are about 20% of the original projectile's size, while most of the fragments are much smaller. This is important, as those ejecta might lead to further collisions with other bodies under similar conditions. As their size is reduced significantly by the first impact, growth will be more probable in a secondary collision.

**Summary and Conclusions**

We present here a first experimental study examining the influence of the impact angle on collisions between dust aggregates at velocities of several tens of m/s relevant for planetesimal formation. The experiments clearly show that large (decimeter or larger) dust aggregates can still gain mass by accreting small particles at collision speeds of 20 m/s and impact angles at least up to 45°. According to the experiments a growth scenario for planetesimals based on following processes is a viable option:

- Decimeter bodies drift through the protoplanetary disk and accumulate mass if the impacting particles are small enough.
- Erosion occurs for particles, which are too large to be accreted directly, but the mass of the target body hardly changes, as erosion is only minor in comparison to the target size. The projectile fragments and target fragments are at a maximum about an order of magnitude smaller in size than the original projectile.
- Ejecta from earlier collisions are accreted by other large aggregates in following collisions.

A related growth scenario has been proposed by Johansen et al. (2008). They assumed threshold velocities depending on the particle sizes, with fragmentation and erosion above that limit and growth below this limit. With the more complex collision results obtained by experiments here, the growth efficiency might be reduced compared to the growth model by Johansen et al. (2008). However, net growth is still possible in a cascade of a small number of projectile-grinding collisions the high velocity range as suggested by Teiser and Wurm (2009a) on the basis of central collisions. If only smaller grains form in collisions, they are small enough eventually, that larger bodies could sweep them up.

Acknowledgements

This work was funded by the Deutsche Forschungsgemeinschaft as part of the research group FOR 759. Our sincere thanks go to Marc Cintala for a thorough review of the manuscript. We also thank the 2nd anonymous reviewer.